# Macrospin analysis of RF excitations within fully perpendicular magnetic tunnel junctions with second order easy-axis magnetic anisotropy contribution

Alexandru Atitoaie[1], Ioana Firastrau[1], Liliana D. Buda-Prejbeanu[2], Ursula Ebels[2], and Marius Volmer[1]

[1]Transilvania University of Brasov, 500036, Brasov, Romania

[2]Univ. Grenoble Alpes, CEA, CNRS, Grenoble INP*, INAC, SPINTEC, F-38000 Grenoble, France

## Abstract

The conditions of field and voltage for inducing steady state excitations in fully perpendicular magnetic tunnel junctions (pMTJs), adapted for memory applications, were numerically investigated by the resolution of the Landau-Lifshitz Gilbert equation in the macrospin approach. Both damping-like and the field-like spin transfer torque terms were taken into account in the simulations, as well as the contribution of the second order uniaxial anisotropy term ($K_2$), which has been recently revealed in MgO-based pMTJs. An in-plane applied magnetic field balances the out of plane symmetry of the pMTJ and allows the signal detection. Using this model, we assessed the states of the free layer magnetization as a function of strength of $K_2$ and polar $\theta_H$ angle of the applied field (varied from $90^0$ to $60^0$). There are two stable states, with the magnetization in-plane or out of plane of the layer, and two dynamic states with self-sustained oscillations, called in-plane precession state (IPP) or out of plane precession state (OPP). The IPP mode, with oscillation frequencies up to 7 GHz, appears only for positive voltages if $\theta_H =90^0$. However, it shows a more complex distribution when the field is slightly tilted out of plane. The OPP mode is excited only if $K_2$ is considered and reaches a maximum oscillation frequency of 15 GHz. Large areas of dynamic states with high frequencies are obtained for strong values of the field-like torque and $K_2$, when applying a slightly tilted external field toward the out of plane direction. The non-zero temperature does not modify the phase diagrams, but reduces drastically the power spectral density peak amplitudes.

# I. INTRODUCTION

In a standard CPP magneto-resistive (MR) structure, the spin angular momentum of the conduction electrons, which are spin polarized in the direction of the pinned layer magnetization, is transferred to the free layer magnetization, yielding a torque, called spin transfer torque (STT) [1,2]. Via STT, the magnetization of the free layer can be switched, with applications in magnetic random access memories (STT–MRAMs) [3], or can be driven into steady state oscillations, with applications in spin transfer nano-oscillators (STNOs) [4]. Until now, the characteristics of MR cells have been optimised in order to fulfil independently the requirements of MRAMs or STNOs. However, the use of the same stack for both applications is imposing as the suitable solution to the increasing demand of energetic efficient miniaturized devices with low fabrication costs from the last decade. Recent studies have shown that fully perpendicular magnetic tunnel junctions (pMTJs) based on materials with strong interfacial perpendicular anisotropy [5, 6] provide better thermal stability at reduced dimensions. They also provide smaller size and lower critical current for current-induced magnetization switching [7-9] than the planar configuration, currently emerging as best candidates to develop high performance STNOs or STT-MRAMs. Nevertheless, in these all perpendicular MR structures, triggering the magnetization precession or generating a switching event are possible only if an initial non-zero torque is produced. Generally, this role is played by thermal fluctuations [10, 11] and has a stochastic character. This difficulty can be overtaken by considering an additional in-plane shape anisotropy [12], by applying an in-plane magnetic field or, lately, by using materials with relevant values for both first and second order anisotropy energy densities [13-16]. These solutions have been mostly investigated for STT-MRAMs, while the requirements for inducing steady state oscillations have not been determined and fully understood yet. Therefore, we propose in this paper to numerically investigate the conditions necessary to excite self-sustained magnetization oscillations in pMTJs subject to an in-plane (or slightly out of plane) applied magnetic field, when taking into account the second order anisotropy term effect. The study is carried out for parameters corresponding to state-of-the art pMTJ structures, used for memory applications. The goal is to theoretically explore whether the same MR stack could provide both memory and communication capabilities.

The paper is organized as follows: after a brief description of the numerical model used in simulations (Section II), within the Section III, we are focusing on the determination of ranges for the applied magnetic field and voltage that are necessary to observe steady state oscillations, when taking or not into account the second order anisotropy term. A more detailed analysis of the second order anisotropy influence on the FL magnetization dynamics is carried out in Section IV. Section V of the paper concerns the outcome of a slightly out of plane tilted applied magnetic field, pointing out new wider ranges for voltage and field for which steady state oscillations occur. The influence of the temperature on the voltage-field diagrams and amplitude/frequency characteristics is also presented in Section VI. Section VII summarises the conclusions of this numerical study.

## II. NUMERICAL MODEL

A schematic representation of the pMTJ that we model is illustrated in Fig. 1(a). The pinned layer magnetization is assumed to be fixed and oriented parallel to the negative out of plane direction (-OZ). A current injected into the structure is spin polarized in the direction of the pinned layer magnetization, which is consequently called polarizer (POL), and creates a torque through the spin-transfer effect [1,2] on the free layer (FL) magnetization. The dynamic motion of the FL magnetization vector $\mathbf{M}(t)$ is given by the phenomenological Landau-Lifshitz-Gilbert (LLG) equation [17, 18], which contains the precession term and the damping term (first two terms of eq.

1 presented here in the Gilbert form). The torque developed by the spin polarized current (STT), creates effects that in MTJs are described by adding two new terms to the Gilbert equation: the damping-like spin transfer torque (3rd term of eq. 1) and the field-like spin transfer torque term (4th term of eq. 1). With this the full equation reads:

$$\frac{d\mathbf{m}}{dt} = -\gamma(\mathbf{m} \times \mu_0 \mathbf{H}_{\text{eff}}) + \alpha\left(\mathbf{m} \times \frac{d\mathbf{m}}{dt}\right) - \gamma a_{//} V[\mathbf{m} \times (\mathbf{m} \times \mathbf{p})] + \gamma a_{\perp} V^2 (\mathbf{m} \times \mathbf{p}), \quad (1)$$

where $\mathbf{m}(t) = \frac{\mathbf{M}(t)}{M_S}$ is the normalized magnetization vector of the FL, with $M_S$ the saturation magnetization, $\gamma$ is representing the gyromagnetic factor of free electrons, $\mu_0$ is the vacuum permeability, and $\alpha$ is the Gilbert damping constant, describing the relaxation in the ferromagnetic system. $\mathbf{p}$ is the electrons spin-polarization vector, which is parallel to the POL magnetization. $a_{//}$ represents the amplitude of the damping-like torque term, $\left(\frac{d\mathbf{m}}{dt}\right)_{damping-like} = -\gamma a_{//} V \mathbf{m} \times (\mathbf{m} \times \mathbf{p})$, which is a term proportional to voltage V, while $a_{\perp}$ is the amplitude of the field-like torque term, $\left(\frac{d\mathbf{m}}{dt}\right)_{field-like} = \gamma a_{\perp} V^2 (\mathbf{m} \times \mathbf{p})$, proportional to $V^2$ for symmetric MTJs [19,20]. The field-like torque is generally neglected in metallic nanostructures, but in MTJs has to be taken into account and could reach an amplitude comparable to that of damping-like torque, as estimated in different theoretical and experimental papers [12, 19-22]. However, it should be mentioned that the dependence of the field-like torque on the bias voltage still needs to be clarified, as different papers report both quadratic [19-21] and linear [23-25] variation on V. $\mathbf{H}_{\text{eff}}$ denotes the effective field that in the macrospin approximation accounts for the magnetic anisotropy field $\mathbf{H_u}$, the externally applied magnetic field $\mathbf{H_{app}}$, and the demagnetizing field $\mathbf{H_D}$. The demagnetizing field $\mathbf{H_D}$ is estimated using the demagnetizing tensor $\overline{\overline{N}} = (N_{xx}, N_{yy}, N_{zz})$ calculated by supposing a rectangular shape of the magnetic layers [26]. The applied magnetic field **H**$_{app}$ is parallel to the OX direction or has a small out of plane component, tilted away from OZ axis with the angle $\theta_H$ (see Fig. 1). The asymmetry produced by the in-plane applied field leads to a non-zero variation of the dynamic magnetoresistance that is given by the projection of the FL magnetization **m** onto the POL axis **p**. The corresponding voltage change enables the detection of the STT induced oscillations. Moreover, the first theoretical predictions [27] and very recent experiments [13-16] show that a second order easy-axis magnetic anisotropy term has to be taken into account in thin perpendicularly magnetized layers that constitute MgO-based pMTJs. It is considered that it arises from spatial fluctuations of the film thickness that generates a crystallographic mismatch at the MgO-magnetic layer interface [13]. For MRAMs, it is expected that this leads to an apparent increase in the spin-torque switching efficiency [8] by reducing the critical current. Here we investigate its impact on the steady state oscillations of a pMTJ. The existence of this higher order term modifies the anisotropy contribution to the free energy density as follows:

$$E_k = -K_1(\mathbf{u}_k \cdot \mathbf{m})^2 - K_2(\mathbf{u}_k \cdot \mathbf{m})^4, \tag{2}$$

where $K_1$ and $K_2$ are constants of the first and second order anisotropy energy, and $\mathbf{u}_k$ is the uniaxial anisotropy direction perpendicular to the film surface (Oz). For specific conditions $K_1 - \mu_0(N_{zz} - N_{yy})M_s^2 \cos^2\theta_c / 2 + 2K_2 < 0$ [16]) the ground state of the magnetic system evolves from the easy-axis state, parallel to the normal of the layers, to a new state called easy-cone state. In the easy-cone regime, the magnetization is tilted away from the out of plane direction, the system energy remaining invariant around a cone with a certain opening angle $\theta_c$ (see Fig. 1(b) that shows the out of plane component of the FL magnetization m$_z$ at equilibrium as a function of K$_2$ in zero voltage and zero applied field). Using this expression for the magnetic anisotropy, we numerically calculated the voltage-field static and dynamic diagram of states of the FL magnetization by integrating, for each value of field and voltage, the modified LLG equation. In this study, the variation of the perpendicular magnetic anisotropy, due to the electric field developing across the tunnelling barrier when applying a dc voltage [28-31] is neglected. Moreover, we assumed that only the FL magnetization can evolve under the action of the effective field and of the perpendicularly injected current, while the magnetization of the POL is fixed. The physical and geometrical parameters used in the simulations are summarized in Table I.

TABLE I. Parameters used in simulations.

| Parameters | Free Layer |
|---|---|
| side dimensions (nm×nm) | 144×144 |
| thickness (nm) | 1.0 |
| $M_S$ (kA/m) | 1000 |
| $K_1$ (kJ/m$^3$) | 638.3 // OZ |
| $K_2$ (kJ/m$^3$) | 0, -10, -30, -60, -120 |
| $N_{xx}, N_{yy}, N_{zz}$ | 0, 0, 1 |
| $\alpha$ | 0.01 |
| $a_{||}$ (mT/V) | 12 |
| $a_{\perp}$ (mT/V$^2$) | 24 |
| $(p_x, p_y, p_z)$ | (0, 0, -1) |

For the numerical integration we applied a predictor-corrector Heun scheme, taking into consideration an integration time step of 1 fs, a delay between voltage step and field of 200ns, and a rise time of voltage step of 50ps.

## III. VOLTAGE-FIELD DIAGRAMS OF STATES FOR pMTJS WITH IN-PLANE APPLIED MAGNETIC FIELD

First, we analysed the static and dynamic magnetization states of FL within a pMTJ in the presence of an in-plane applied magnetic field. As already mentioned, the use of an in-plane field is motivated by the fact that an asymmetry needs to be produced between the FL oscillation and the POL in order to induce a non-zero variation of the dynamic magneto-resistance, and to allow the signal detection. We performed two sets of simulations: one set that does not consider the second order anisotropy contribution ($K_2=0$) and one that takes into account a high value for the $K_2$ constant ($K_2$=-120 kJ/m$^3$). Before analysing the dynamic excitation spectra under a spin polarized current, we first simulated the static equilibrium states of the FL magnetization as a function of the magnetic field applied at $\theta_H = 90^0$. For material parameters listed above (Table I), the hard-axis hysteresis loop of the FL exhibits a saturation field of $\mu_0H_{sat}$=20mT. When the spin polarised current effects are taken into account, different static and dynamic magnetization states are obtained as a result of the competition between the torques acting on it. These states can be well illustrated on a voltage-field state diagram such as that presented in Fig. 2(a) for the case $K_2$=0. The colour code shows the out of plane $m_z$ component of the FL magnetization, which ranges between the red colour, specific for the positive orientation of FL magnetization as against the OZ axis, and toward blue colour, specific for the negative orientation. On this phase diagram one can notice that for low values of voltage and high values of the applied field, the spin-torque, which tends to bring the FL magnetization out of plane, is not strong enough to induce any dynamics and the magnetization remains in the plane of the layer, parallel to the applied field direction. This state is called in-plane stable state, that is depicted in the diagram with IPS (+/- are showing the parallel respectively antiparallel orientation of the FL magnetization to OX axis). In the case of high voltages the larger spin-torque stabilizes the magnetization out of plane, in a direction parallel to the OZ axis. This state is denoted out of plane stable state (OPS on the diagram, with + or - keeping the same sign convention as before). In between these two stable states the equilibrium state is one with an intermediate angle in the OXZ plane (as sketched by the small black arrows on the diagrams). However, for positive voltage values and for applied fields beyond the saturation field, the FL magnetization passes from OPS- to an in-plane state by crossing a shaped-stripe area in which the magnetization is driven into self-sustained oscillations. This dynamic state is called in-plane precession state (IPP). The corresponding trajectories at constant voltage for three different field values are illustrated in the inset of Fig. 2(b), which gives the frequency range within the voltage-field plane. Here, the colour code shows that the frequency varies in between 1 and 7 GHz in the range of fields and voltages considered in the calculations. The peaks amplitude of the power spectral density (PSD) plot, not shown here, reaches maximum values in the central zone of the IPP stripe (around 0.26 a.u) and minimum values (around 0.10 a.u.) close to its frontiers with the IPS+/- stable states areas. This dynamic region is of interest, since the out of plane component of FL magnetization during one oscillation period has a substantial variation (see inset of Fig. 2(b)), giving rise to a large resistance change and consequently the signal detection is insured. However, for a constant field, the voltage range $\Delta V$ for which the oscillations exist are rather small with $\Delta V \approx 160 mV$. For negative voltages no dynamic IPP state is observed. This asymmetry in voltage of the state diagram can be understood by analysing the individual contribution of each term of STT. In our convention, positive voltage implies electrons flowing from POL to FL. Consequently, for positive/negative voltage the damping-like STT favours the parallel/antiparallel state of the MTJ, while the field-like STT favours the antiparallel state regardless of the voltage sign. This is deduced from Fig. 2(c), where phase diagrams are compared when only the damping-like ($a_\parallel \neq 0$ and $a_\perp = 0$) or only the field-like ($a_\perp \neq 0$ and $a_\parallel = 0$) torque is present. If the voltage is positive, the damping-like torque favours the parallel state of MTJ (OPS-), while the field-like torque favours the antiparallel state (OPS+). The competition between these two spin transfer torques results in the dynamic equilibrium IPP state of Fig. 2(b). However, this is not the case for negative voltage, where both damping-like and field-like torques favour the antiparallel state (OPS+) and, consequently, there are no oscillations. The slightly curved appearance of the IPP stripe is related to the square dependence on V of the field-like torque [12].

The shape of the phase diagram changes when the second order anisotropy term is taken into account. The static and dynamic voltage-field diagrams calculated for $K_2 = -120$ kJ/m$^3$ are depicted in Fig. 2(d) respectively in Fig. 2(e). One can observe the same three states as in the previous case, namely the stable states IPS+/-, OPS+/-, and the dynamic IPP state. However the IPS state completely disappears for positive voltages beyond the IPP band (in the range of voltage considered here), and the dynamic IPP state areas are significantly larger for both positive and negative directions of the applied magnetic field. The voltage range $\Delta V$ of the IPP stripe at a constant field value doubles compared to the case of $K_2 = 0$. Furthermore, an additional dynamic state is appearing in a triangular shaped zone around $\mu_0 H_{app} = 0$, and for both signs of voltage. Analysing the magnetization trajectories for these oscillations as shown in the inset of Fig. 2(e), one can observe that they are taking place around an out of plane direction and at frequencies ranging from 1 GHz to a maximum of 14 GHz, double than in the previous situation. While the oscillation frequency increases with increasing voltage, the dependence in field is not important (decreases slightly with increasing field), because the range of fields where this dynamic state is observed is quite small. This behaviour in voltage and magnetic field alongside to the triangular shape of these oscillation zones are features of the out of plane precession states (OPP), similar to that obtained in a spin-torque oscillator with perpendicular POL and in-plane magnetized FL, as has been proofed by analytical calculations [32], numerical simulations [33], and experimental measurements [34]. In fact, at large values of $K_2$ the ground-state of the FL magnetization is the easy-cone state with the polar angle of the cone close to 90° (see the Fig. 1(b)) that mimics the conditions of an in-plane uniaxial magneto-crystalline anisotropy as considered in the previously cited papers. The important advantages from the applications point of view are that the OPP mode is obtained in zero applied field and reaches quite high oscillation frequencies. However, the weak point is that the amplitude of these oscillations and consequently the MR signal during one period of oscillation are quite small, making these oscillations very difficult to detect in the absence of an additional analyzer.

## IV. SECOND ORDER ANISOTROPY EFFECTS

In order to better understand the effects of the second order anisotropy contribution, we investigated the influence of the strength of $K_2$ on the voltage range $\Delta V$ and on the frequency of IPP/OPP oscillations. Therefore, we calculated the voltage-field diagrams for three values of the second order anisotropy that are weaker than in Figs. 2(d, e), namely $K_2$=-10, -30, -60 kJ/m$^3$. The $K_2$ threshold for which the FL magnetization passes from the easy-axis (parallel to OZ) to the easy-cone state is $K_2$=-4.9 kJ/m$^3$ [16], indicating that for the $K_2$ values considered above the FL is in the easy-cone ground state (see the Fig. 1(b)). Figs. 3(a, b, c) display the simulations results for the first quadrant (positive voltage and positive applied field) of phase diagrams, the colour code showing, by reading from up to down, first line, the in-plane ($m_x$) component of FL magnetization, second line, the out of plane ($m_z$) component of FL magnetization, and third line, the oscillation frequency. As already pointed out, the IPP oscillations region appears mostly as a 'switching' band between the two in-plane stable states, IPS+ and IPS-. However, with increasing $K_2$ value, and for weaker applied fields this dynamic IPP zone marks also the transition from the IPS+ state directly to the OPS- state. The maximum field value for which the FL magnetization passes from an in-plane state directly to an out of plane one is proportional to $|K_2|$. Additionally, the overall area of IPP mode and consequently the width $\Delta V$ of the oscillation band is generaly increasing by increasing the second order anisotropy term. This behaviour of IPP oscillations is better depicted in Fig. 3(d), in which the frequency values for different $K_2$ values are plotted against the voltage for an external applied field maintained constant. It is shown that the highest voltage range of 0.5V is obtained for $K_2$=-120 kJ/m$^3$, while the voltage range drops by a third for $K_2$=-60 kJ/m$^3$ and by one half for $K_2$=-30 kJ/m$^3$ or lower, respectively. In all cases the frequency exhibits a linear dependence on voltage, increasing with increasing voltage. The PSD peak amplitude has a maximum for voltages corresponding to the middle of the IPP band if $|K_2|$ <30 kJ/m$^3$, and increases with V for large $|K_2|$ (see Fig. 3(e)). This behaviour is due to the fact that, for low $|K_2|$, the FL magnetization evolves from an in-plane precession around the +Ox axis to an in-plane precession around the -Ox axis (see schematics at the bottom of Fig. 3). During this process, the opening of the magnetization trajectory first increases with voltage V and after decreases, with a maximum in the centre of the IPP band that results in a maximum of PSD peak amplitude. On the contrary, for large $|K_2|$, the magnetization precession occurs only around +Ox axis, with a trajectory opening that increases with increasing V. Accordingly, the PSD amplitude is increasing too. The frequency is also blue-shifted upon magnetic field as illustrated in Fig. 3(f), that are representing the frequency variation at constant voltage (V=0.5V). The PSD peak amplitude generally shows the same evolution as presented before, as a function of $|K_2|$ strength (see Fig. 3 (g)). Nevertheless, for large $|K_2|$, the FL magnetization, which is out of plane at very low fields (OPS-), starts to describe steady state precessions on a large in-plane trajectory when the external magnetic field reaches a critical value. If the field is further increased, it tends to bring the magnetization in-plane and the trajectory opening is reduced (see draws at bottom of Fig. 3). The PSD peak amplitude reduces as well. All these graphs show that the effect of the second order anisotropy on the IPP oscillation frequency is pronounced for large $|K_2|$, leading to a significant enlargement of the IPP zone as compared to the cases where $|K_2|$ <60 kJ/m$^3$ for which the curves almost superpose. The frequency oscillation of the OPP mode has a 'classical' dependence on voltage and field [32] with the frequency increasing proportional to the out of plane component $m_z$ of the FL magnetization. The plots of Fig. 3(h) illustrate the linear behaviour of frequency versus voltage at zero applied field and considering the same $K_2$ variation. Contrary to the frequency that is enhanced by the voltage increase, the amplitude of the PSD peak is decreasing with increasing voltage, as shown in the Fig. 3(i). This monotonous drop is due to the shrink of the magnetization trajectory as it goes out more and more out of plane with increasing V (see the inset of Fig. 2(e)).

Hence, these results clearly show that the second order anisotropy term is definitely changing the phase diagram especially for low fields and also assists in reaching larger regions of dynamic states of higher frequency self-sustained oscillations with both in-plane and out of plane precessions.

## V. OUT OF PLANE FIELD ORIENTATION EFFECTS

We also investigated the influence of an external field applied at a tilt angle $\theta_H$ on the characteristics of the self-sustained oscillations of pMTJs. The dynamic voltage-field diagrams were simulated including a second order anisotropy $K_2$, ranging from 0 to -120 kJ/m$^3$ while the angle $\theta_H$ varied between 60° and 80°. The corresponding results are shown in Fig. 4.

The first remark concerns the case $K_2$=0 and represents the fact that the symmetry in field observed in Figs. 2(a, b) for a field applied along the OX axis vanishes even at small deviations from the in-plane orientation (Fig. 4(a)). Analysing the contributions of different terms in the LLG equation, we concluded that for fields having both in-plane and out of plane components the dynamic IPP zone separating the IPS+ and IPS- states appears only in quadrants 1 and 3 as a tilted line (diagram not shown), if the field-like torque is not taken into account (a⊥=0). This shape of diagram with inclined critical lines is similar to those derived analytically in the Ref. [12] for pMTJs, having a two components anisotropy contribution, one in plane (//OX) and one out of plane (//OZ). If the field is applied only in-plane (//OX) and a⊥≠0, the effect of the filed-like torque is similar to the that of a tilted field, but as the filed-like torque is proportional to the voltage square the IPP state appears only for positive voltages, in quadrants 1 and 2, displaying a symmetric distribution against the field sign (as presented in Figs. 2(a, b)). When a tilted field and the field-like torque are considered at the same time, the competition between different torques results in the appearance of the dynamic IPP mode in three quadrants, 1, 2, and 3, as presented in Figs. 4(a, d, g). In quadrant 3 (H<0, V<0) this mode shows an interesting loop shape whose size increases as the out of plane deviation of the applied field is increasing (for V and $\mu_0 H_{app}$ ranges considered on the diagrams the loop is only partially visible).

Second, we noticed that, generally, the width of the IPP stripe reduces when the field moves away from the in-plane orientation, suggesting that a field applied normal to the layers surface tends to suppress oscillations, as previously shown in Ref. [12]. Moreover, for the positive values of the field and voltage, the slope of the IPP stripe increases proportionally with $\theta_H$. However, the frequency values seem to depend only on the total field amplitude and not on the field orientation (for example for $\mu_0 H_{app}$=0.1T the IPP frequency ranges between 1.75 and 3.5 GHz for all angles $\theta_H$ taken into account). Further, when the second order anisotropy term is considered in the simulations, the phase diagrams generally appear as a combination between those obtained for $K_2$=0 and the diagram of Fig. 2(e) (where $K_2$=-120 kJ/m$^3$ and the field is in-plane). The main contribution of $K_2$ is the appearance around the zero-field line of a triangular zone with out of plane oscillations (OPP), that are having higher frequencies and larger range in the V-H diagram for stronger $K_2$ values (as it was shown above). The second order anisotropy influence on those oscillations is not (or very little) affected by the angle of the external field. Nevertheless, a larger $K_2$ may increase the width of the IPP and OPP mode zone, as can be seen in Figs. 4(c, f, i).

Finally, the most significant result that emerges from these calculations is that there is a combination of $\theta_H$ and $K_2$ for which, for negative values of both field and current, the IPP mode exists in a very large range of H and V values. This case is illustrated on the diagram of Fig. 4(f), which is obtained for the angle $\theta_H$ =70° and $K_2$=-120 kJ/m$^3$. Here, the IPP zone reaches a width of 150mT in field and 600mV in voltage, which is very convenient for STNOs applications, even if the PSD amplitude is approximatively a third of that estimated for positive voltages.

## VI. TEMPERATURE EFFECTS

The simulations summarized above do not accounting for the effect of the thermal fluctuations, the sample being at 0K. We have also analysed the effect of the temperature on the FL magnetization states and the characteristics of self-sustained oscillations for the situation where $K_2=0$ and $\theta_H=90^0$. The effect of temperature was included in the macrospin model in a similar way as in Ref. [35]. Accordingly, we simulated the dynamic voltage-field diagrams for a system exposed to temperatures of 50, 100 and 300K. The results corresponding to the first quadrant (V>0, H>0) are shown in Figs. 5(a, b, c), where the colour code is designating the power spectral density peak amplitude. We have also displayed in Figs. 5(d, e, f) the PSD vs. frequency for several values of voltage and the same value of the field. From these graphs, we can conclude that, even though the shape and the width of the IPP mode band are not affected by temperature, the PSD peak amplitude decreases considerably when the system is heated. More precisely, for the same voltage and field values, the centre of the spectra does not shift with temperature, but the PSD peak for 300K is one order of magnitude smaller than the one for 0K. Similarly, the triangular zone of the OPP mode is not affected by the addition of temperature (phase diagrams are not shown here), but, as expected, the PSD amplitude knows a drop as temperature grows from 0 to 300K. Figs. 5(g, h, i) illustrates this behaviour for $K_2=-120$ kJ/m$^3$ and $\mu_0H_{app}=0.001$T. Therefore, we can state that the H-V range of the in-plane and out of plane precession states are not affected by the temperature, but the amplitudes are decreasing drastically at room temperature.

## VII. CONCLUSIONS

We have analysed the FL magnetization states of a fully perpendicular magnetic tunnel junction using the LLG equation within a macrospin approach, considering both damping-like and field-like spin transfer torques and, also, taking into account the second order anisotropy term. We showed that, when the damping-like and field-like spin transfer torques are considered, the voltage-field diagrams present one out of plane stable state (OPS+/-), one in-plane stable state (IPS+/-) and two dynamic areas with in-plane precession states (IPP), with frequencies ranging from 1 to 7 GHz.

When the second order anisotropy term is introduced, a new dynamic state with out of plane oscillations (OPP) is obtained for lower values of external field and induced current with frequencies up to 14 GHz. The effect of temperature is discussed, as well as the influence of a tilted external field toward the out of plane direction on the free layer magnetization states, showing the existence of a large area of in-plane precession state for negative values of field and voltage only when the external field is tilted.

In conclusion, for larger areas of dynamic states with higher frequencies that can be used for RF applications of STNO that are based on fully perpendicular magnetic tunnel junctions, one must consider stronger values of the field-like torque and of the second order anisotropy term and, also, to use a slightly tilted external field toward the out of plane direction.

**Acknowledgements**: Funding from the European Union's Horizon 2020 research and innovation programme under grant agreement No 687973, acronym GREAT, is highly acknowledged.

FIG. 1. (a) Schematics of the perpendicular MTJ structure; (b) Evolution of the out of plane component of the FL magnetization function of the second order anisotropy constant, $K_2$ value, in zero applied magnetic field and zero voltage.

FIG. 2. (a) Static (out of plane $m_z$ component of FL magnetization) and (b) dynamic (oscillation frequency) voltage-field diagram of states of the FL magnetization without taking into account the second order anisotropy term, $K_2=0$; (c) Comparative study concerning the contribution of parallel ($a_{||}$) and perpendicular ($a_\perp$) spin transfer torques on the static voltage-field diagram of states at $K_2=0$; (d) Static and (e) dynamic voltage-field diagram of states of the FL magnetization by taking into account the second order anisotropy term contribution, $K_2=-120$ kJ/m$^3$. In all cases $\theta_H=90°$.

FIG. 3. Static and dynamic voltage-field diagram of states of the FL magnetization for the first quadrant for (a) $K_2=-10$ kJ/m$^3$, (b) $K_2=-30$ kJ/m$^3$, and (c) $K_2=-60$ kJ/m$^3$. For clarity, both $m_z$ (// out of plane OZ direction) and $m_x$ (// to magnetic field in-plane OX direction) components change are illustrated; (d) IPP frequency, and (e) power spectral density (PSD) peak amplitude vs. voltage evolution at constant field, $\mu_0 H_{app}=0.05$T, for different $K_2$ values; (f) IPP frequency, and (g) PSD peak amplitude vs. field evolution at constant voltage, V=0.5V, for different $K_2$ values; (h) OPP frequency, and (i) PSD peak amplitude vs. voltage evolution at zero applied field for different $K_2$ values.

FIG. 4. Dynamic voltage-field diagrams of states of the FL magnetization function of $K_2$ strength and polar $\theta_H$ angle of the applied magnetic field for (a) $K_2=0$ and $\theta_H=80°$, (b) $K_2=-60$ kJ/m$^3$ and $\theta_H=80°$, (c) $K_2=-120$ kJ/m$^3$ and $\theta_H=80°$, (d) $K_2=0$ and $\theta_H=70°$, (e) $K_2=-60$ kJ/m$^3$ and $\theta_H=70°$, (f)

$K_2$=-120 kJ/m³ and $\theta_H$=70°, (g) $K_2$=0 and $\theta_H$=60°, (h) $K_2$=-60 kJ/m³ and $\theta_H$=60°, (i) $K_2$=-120 kJ/m³ and $\theta_H$=60°.

FIG. 5. Power spectral density (PSD) peak amplitude representation as a function of voltage and field at (a) T=50K, (b) T=100K, and (c) T=300K. PSD representation at constant field ($\mu_0 H_{app}$=0.3T) and $K_2$=0 (IPP mode) for different voltage values at (d) T=50K, (e) T=100K, and (f) T=300K. PSD representation at constant field ($\mu_0 H_{app}$= 0.001T) and $K_2$=-120 kJ/m³ (OPP mode) for different voltage values at (g) T=50K, (h) T=100K, and (i) T=300K.

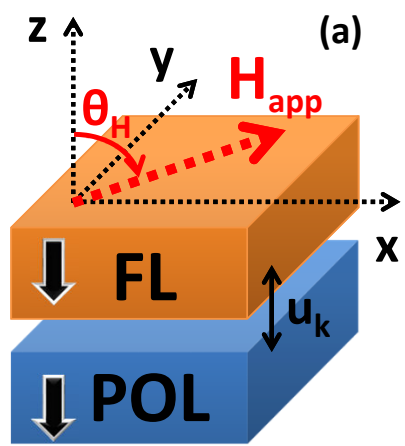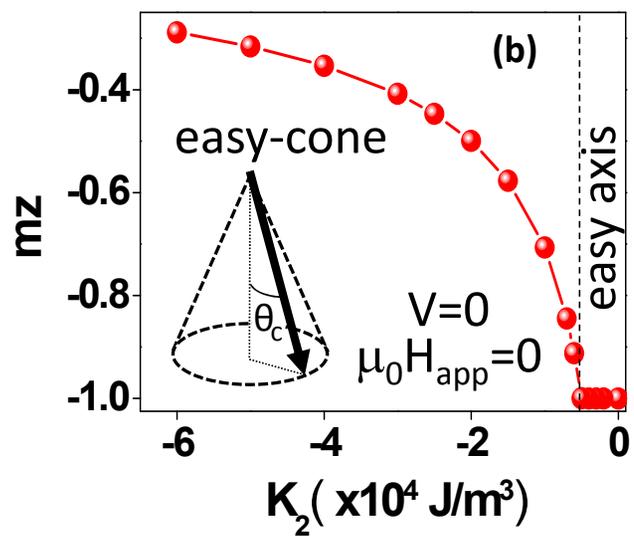

FIG. 1. (a) Schematics of the perpendicular MTJ structure; (b) Evolution of the out of plane component of the FL magnetization function of the second order anisotropy constant, $K_2$ value, in zero applied magnetic field and zero voltage.

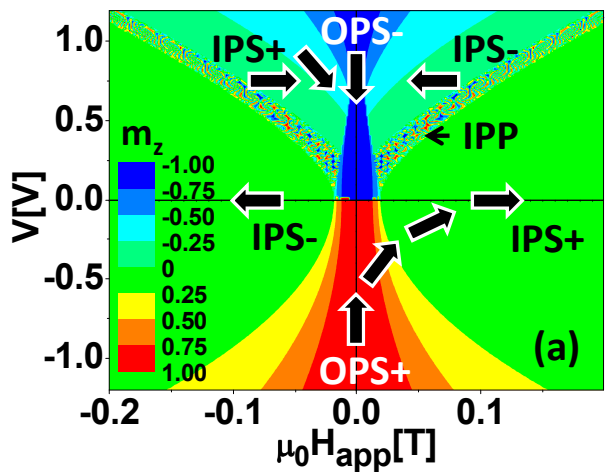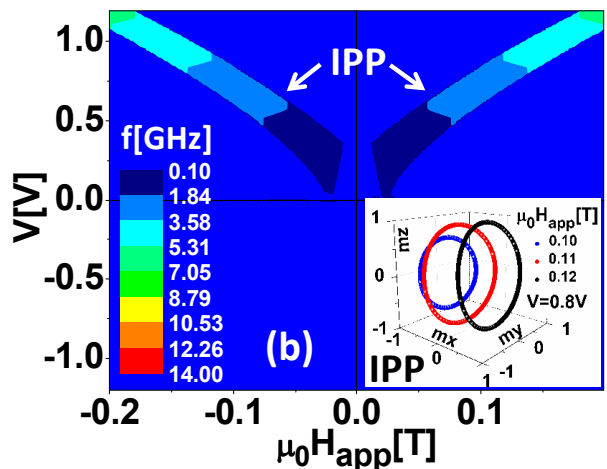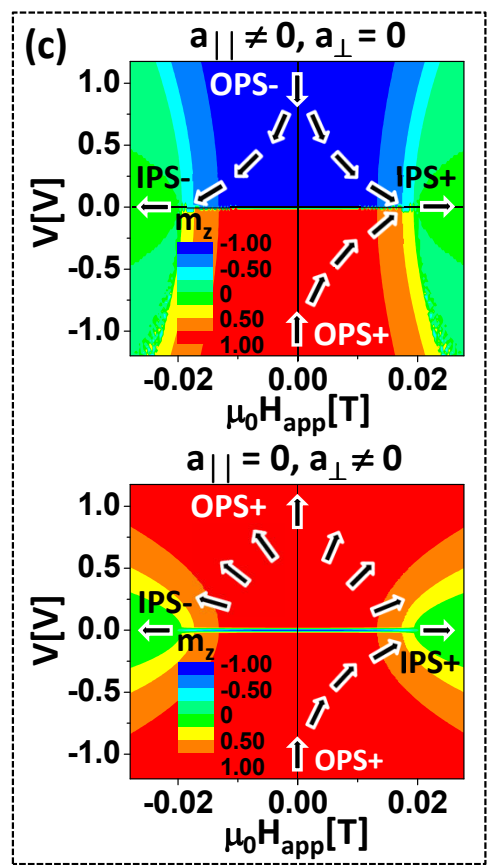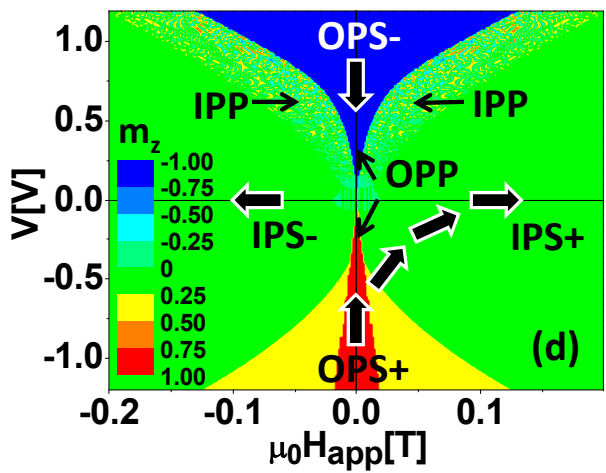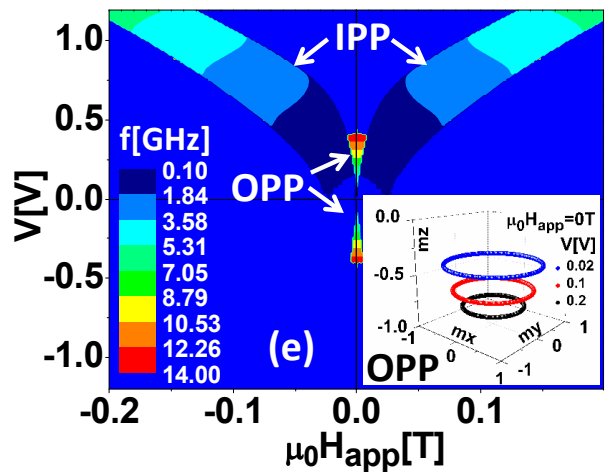

FIG. 2. (a) Static (out of plane $m_z$ component of FL magnetization) and (b) dynamic (oscillation frequency) voltage-field diagram of states of the FL magnetization without taking into account the second order anisotropy term, $K_2=0$; (c) Comparative study concerning the contribution of parallel ($a_{||}$) and perpendicular ($a_\perp$) spin transfer torques on the static voltage-field diagram of states at $K_2=0$; (d) Static and (e) dynamic voltage-field diagram of states of the FL magnetization by taking into account the second order anisotropy term contribution, $K_2=-120$ kJ/m$^3$. In all cases $\theta_H=90°$.

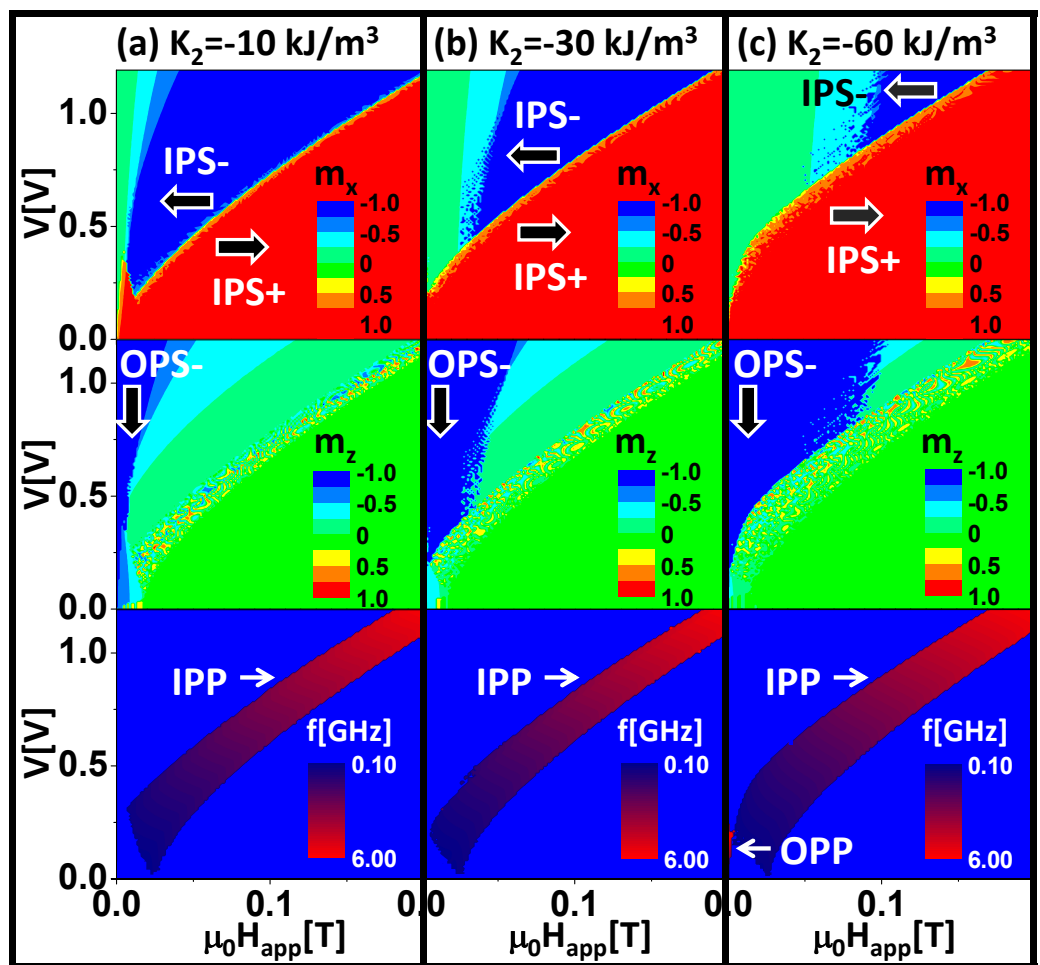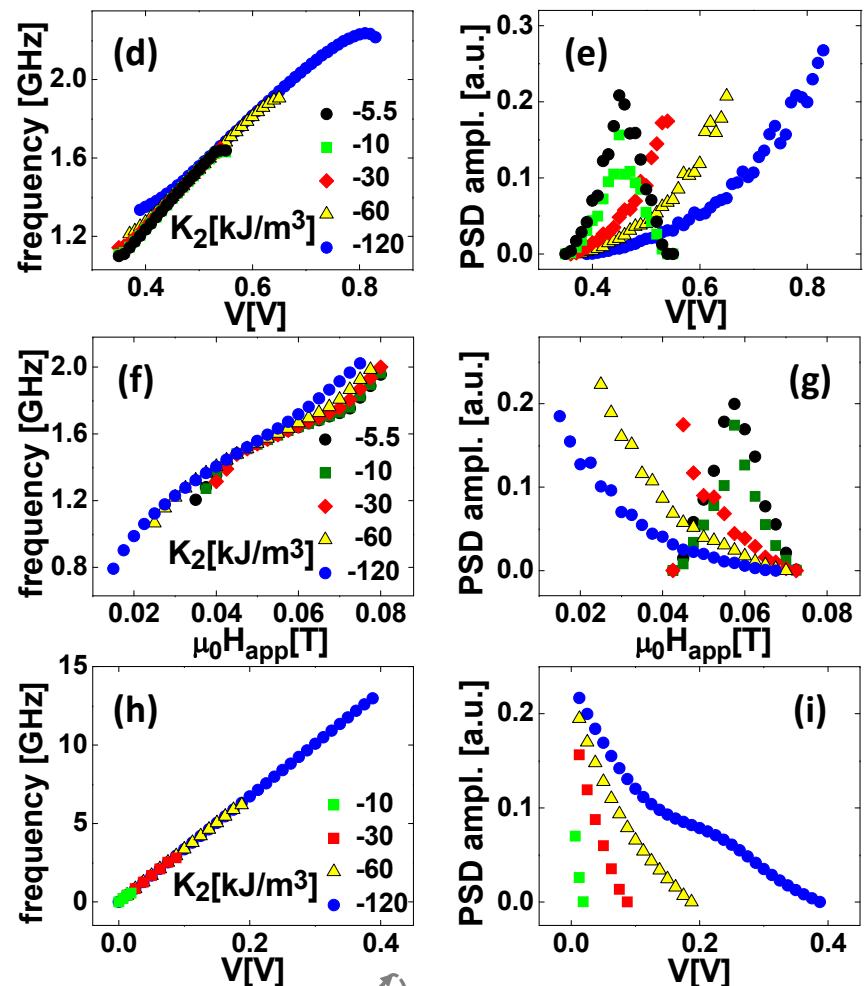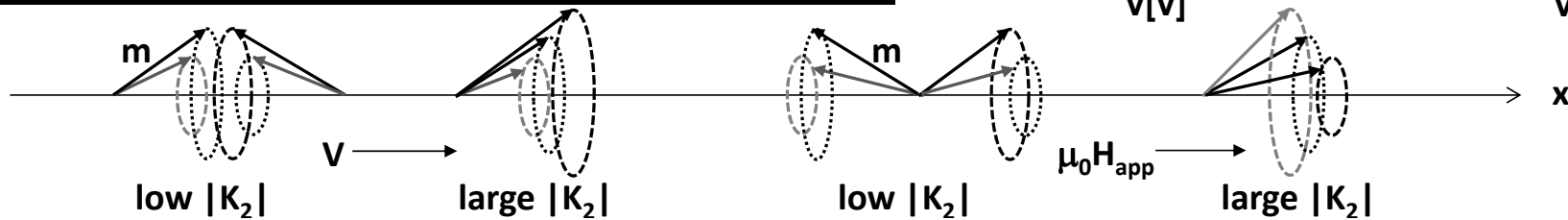

FIG. 3. Static and dynamic voltage-field diagram of states of the FL magnetization for the first quadrant for (a) $K_2$=-10 kJ/m³, (b) $K_2$=-30 kJ/m³, and (c) $K_2$=-60 kJ/m³. For clarity, both $m_z$ (// out of plane OZ direction) and $m_x$ (// to magnetic field in-plane OX direction) components change are illustrated; (d) IPP frequency, and (e) power spectral density (PSD) peak amplitude vs. voltage evolution at constant field, $\mu_0 H_{app}$=0.05T, for different $K_2$ values; (f) IPP frequency, and (g) PSD peak amplitude vs. field evolution at constant voltage, V=0.5V, for different $K_2$ values; (h) OPP frequency, and (i) PSD peak amplitude vs. voltage evolution at zero applied field for different $K_2$ values.

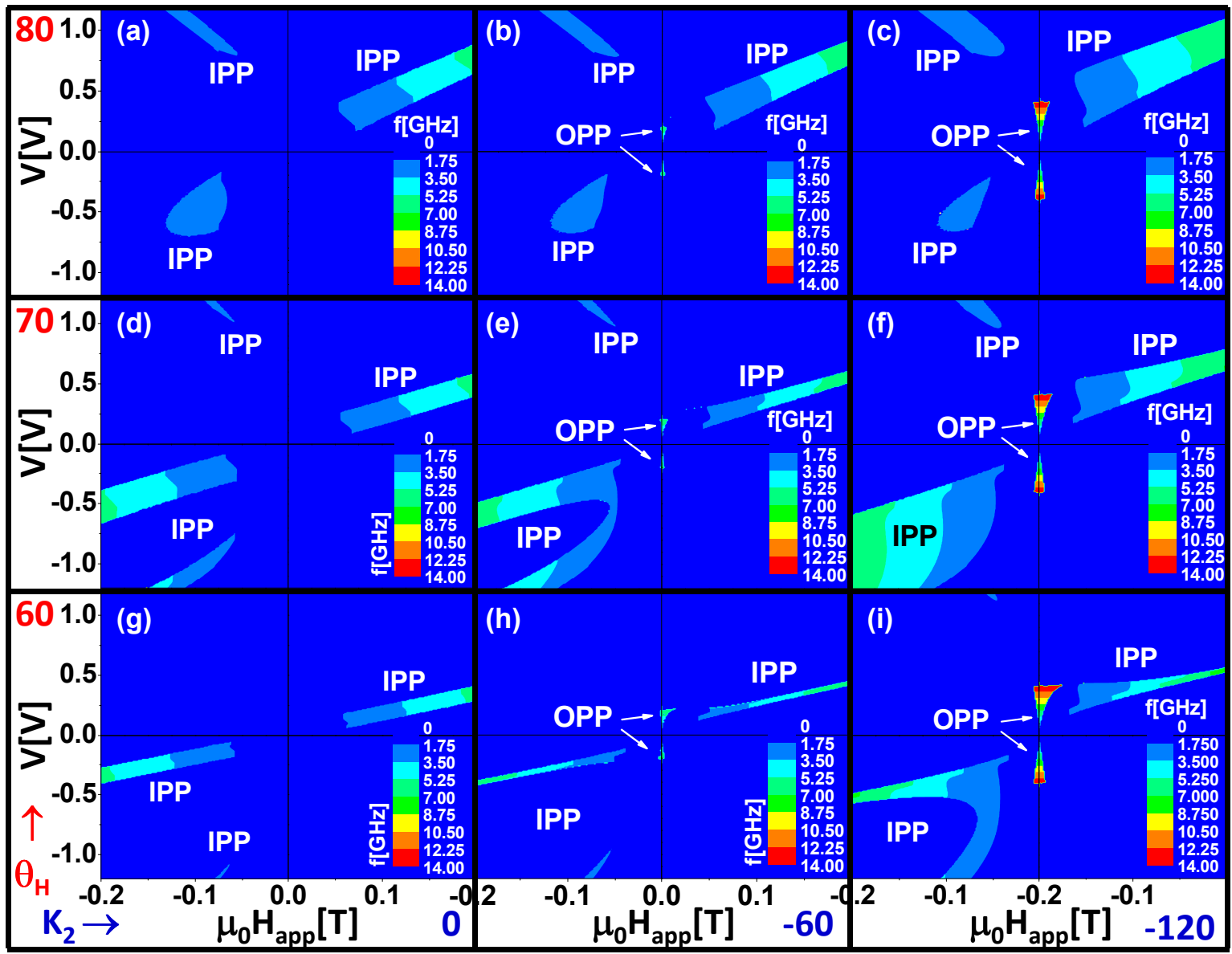

FIG. 4. Dynamic voltage-field diagrams of states of the FL magnetization function of $K_2$ strength and polar $\theta_H$ angle of the applied magnetic field for (a) $K_2=0$ and $\theta_H=80°$, (b) $K_2=-60$ kJ/m$^3$ and $\theta_H=80°$, (c) $K_2=-120$ kJ/m$^3$ and $\theta_H=80°$, (d) $K_2=0$ and $\theta_H=70°$, (e) $K_2=-60$ kJ/m$^3$ and $\theta_H=70°$, (f) $K_2=-120$ kJ/m$^3$ and $\theta_H=70°$, (g) $K_2=0$ and $\theta_H=60°$, (h) $K_2=-60$ kJ/m$^3$ and $\theta_H=60°$, (i) $K_2=-120$ kJ/m$^3$ and $\theta_H=60°$.

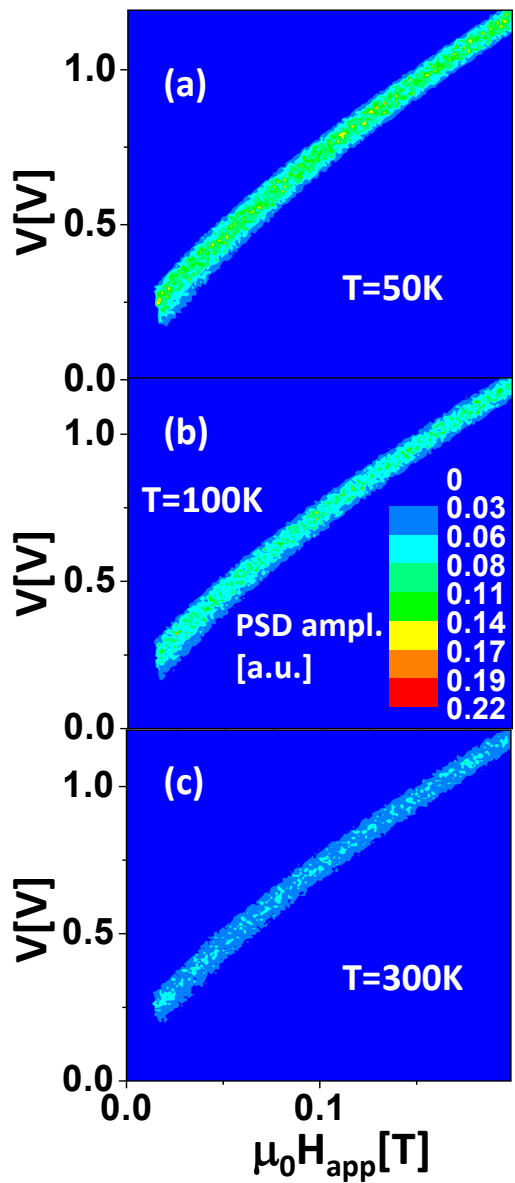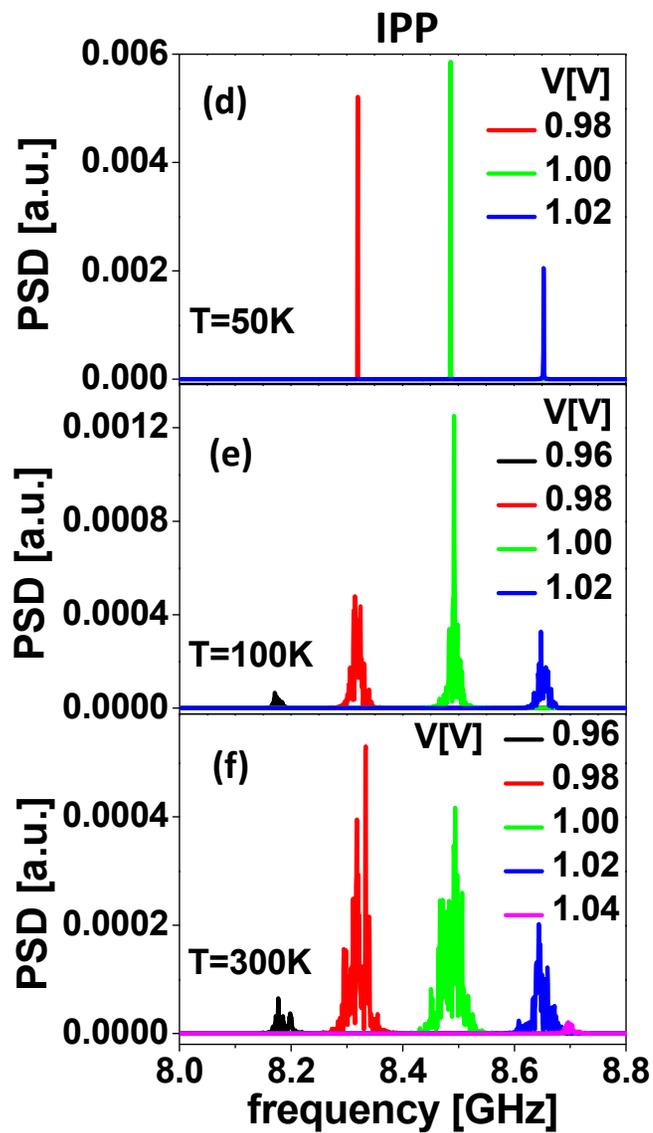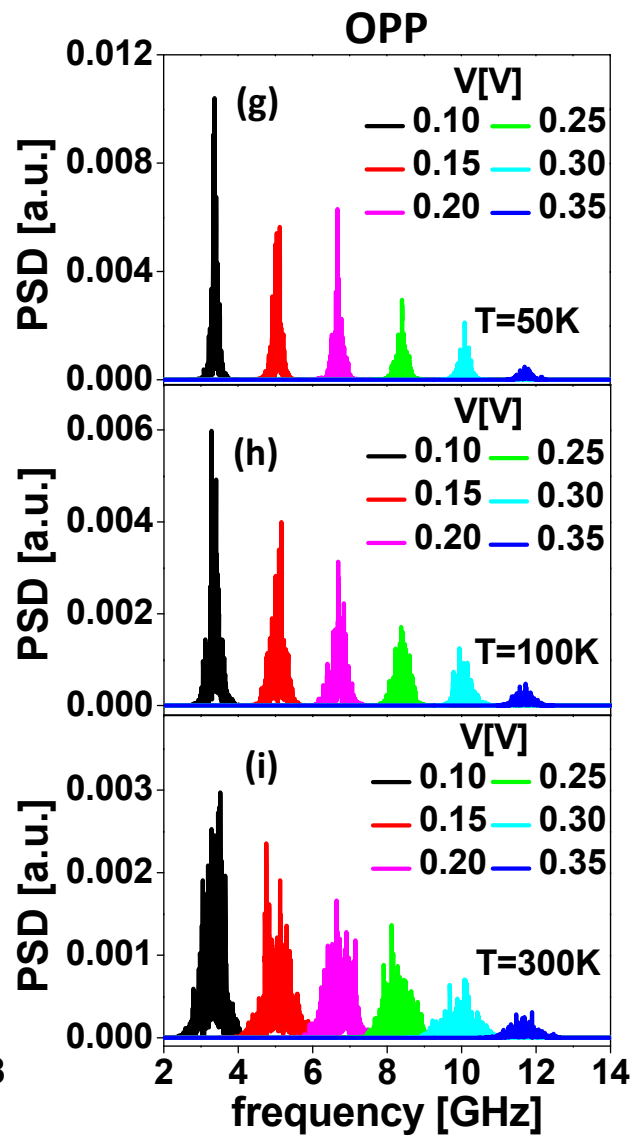

FIG. 5. Power spectral density (PSD) peak amplitude representation as a function of voltage and field at (a) T=50K, (b) T=100K, and (c) T=300K. PSD representation at constant field ($\mu_0 H_{app}$=0.3T) and $K_2$=0 (IPP mode) for different voltage values at (d) T=50K, (e) T=100K, and (f) T=300K. PSD representation at constant field ($\mu_0 H_{app}$= 0.001T) and $K_2$=-120 kJ/m³ (OPP mode) for different voltage values at (g) T=50K, (h) T=100K, and (i) T=300K.